\newcommand{\un}{~\mathrm}
\newcommand{\ie}{{\em i.e. }}
\newcommand{\ea}{{\'e}}
\newcommand{\eg}{{\`e}}

\documentclass[12pt]{iopart}
\usepackage{graphicx}

\begin{document}

\title[Crack fronts and damage in glass at the nanometer scale.]{Crack fronts and damage in glass at the nanometer scale.}

\author{Christian Marli\eg re\dag\footnote[3]{To
whom correspondence should be addressed (christian.marliere@ldv.univ-montp2.fr)}, Silke Prades\ddag, Fabrice C\ea lari\ea \dag, Davy Dalmas\ddag, Daniel Bonamy\dag\ddag, Claude Guillot\ddag ~and Elisabeth Bouchaud\ddag.}  

\address{\dag\ Laboratoire des Verres - UMR CNRS-UM2 5587, Universit\'e Montpellier 2, C.C. 69 - Place Bataillon, F-34095 Montpellier Cedex 5 - France}

\address{\ddag\ Service de Physique et Chimie des Surfaces et Interfaces, DSM/DRECAM/SPCSI, CEA Saclay, F-91191 Gif sur Yvette - France}

\begin{abstract}
We have studied the low speed fracture regime for different glassy materials with variable but controlled length scales of heterogeneity in a carefully mastered surrounding atmosphere. By using optical and atomic force microscopy (AFM) techniques we tracked in real-time the crack tip propagation at the nanometer scale on a wide velocity range ($10^{-3} - 10^{-12}\un{m/s}$ and below). The influence of the heterogeneities on this velocity is presented and discussed. Our experiments reveal also -~for the first time~- that the crack progresses through nucleation, growth and coalescence of {\em nanometric damage} cavities {\em within the amorphous phase}. This may explain the large fluctuations observed in the crack tip velocities for the smallest values. This behaviour is very similar to what is involved, at the micrometric scale, in ductile fracture. The only difference is very likely due to the related length scales (nanometric instead of micrometric). Consequences of such a nano-ductile fracture mode observed at a temperature far below the glass transition temperature, $T_g$, in glass is finally discussed.
\end{abstract}


\submitto{\JPCM}

\maketitle

\section{Introduction}
Despite many works and recent progresses, materials fracture still exhibits many puzzling aspects. Two different observations seem particularly difficult to reconcile:

\begin{itemize}

\item Brittle materials -~whose most common example is glass~- break abruptly, without first deforming, while in ductile materials -~like metallic alloys~- fracture is preceded by large plastic deformations. In this latter case, the crack is observed~\cite{Pineau95} to progress through the coalescence of micrometric damage cavities nucleated from microstructural defects (second phase precipitates, grain boundaries...).

\item However, quantitative studies reveal that both types of fracture surfaces have a very similar morphology. Fracture surfaces have been shown to be self-affine objects~\cite{Mandelbrot84,Bouchaud97} for both brittle and ductile materials. Two self affine regimes
coexist : at small length scales (up to a length $\xi_c$), the roughness exponent $\zeta$  is
close to 0.5, whereas at large length scale (from $\xi_c$ to a larger length $\xi$), $\zeta$  is around 0.8.
These values of $\zeta$  are {\em universal}, but $\xi_c$ as well as $\xi$  depend on the kind of
material considered, ranging from a few nanometres for glass~\cite{Daguier95} to a
few centimetres for concrete and rocks~\cite{Poor92}. It is worthy to note that, for
ductile materials, $\xi_c$  is of the order of the typical size of the damage
cavities when they coalesce~\cite{Pineau95}. 

This strongly suggests that the observed transition between the two self-affine regimes reflects a change from an intra-cavity structure to an inter-cavity one. Such a scenario was first proposed in~\cite{Pineau95,Bouchaud99,Paun}, based on the observation of the roughness of growing cavities before coalescence in an aluminium alloy and  further developed in a recent paper~\cite{Bouchaud02}.

\end{itemize}

How to reconciliate these two observations? A natural supposition is that a ductile-like -~up to now undetected~-  fracture process exists in brittle materials like glass. In this case, the size of the associated damage cavities at coalescence is expected to be equal to the crossover length $\xi_c$, {\em i.e.} of the order of ten nanometers. The need for an experimental observation of such effect was at the genesis of the experiment described in the following. Our new experimental set-up, based on Atomic Force Microscopy (AFM), is described in section 2. Huge velocity fluctuations in the progression of the tip were detected (section 3) and explained by the observation of nanometric cavities developing ahead of the crack tip (section 4). These cavities were proved to be {\em damage} cavities by using two independent methods (section 5). Finally, the possible origins of these cavities are conjectured and the consequences of such ductile fracture mode in glass are discussed (section 6).

\section{Experimental set-up}

The design of our experiments should meet several successive requirements to achieve the goal described above. Those are listed below:

\begin{enumerate}

\item The apparatus should have a very high spatial resolution (a few tenths of nanometers) and work in real space and real time (no post-mortem studies). Moreover, it should deal with electrically non conductive materials. Surface treatments -~which limit the spatial resolution (by metal deposition for instance)~- are also prohibited. The AFM technology appears then to be the ideal tool that meets all these requirements. Due to the nanometric resolution of AFM  we measure crack speeds lower than $10^{-12}\un{m.s}^{-1}$ with a reasonable acquisition time (15 minutes between each acquired image).

\item The main drawback of AFM resides in its recording time, that cannot be set lower than a few minutes without significant loss in the spatial resolution. It forces us to work in the so-called 'stress corrosion' or 'sub-critical growth' regime~\cite{Wiederhorn67,Wiederhorn70,Michalske91}. The crack behaviour is then controlled by the chemically activated processes with the surrounding environment. A rigorous control of atmosphere and temperature is then necessary.

\item The crack progression should be stable and well monitored. This leads us to design a mechanical DCDC (Double Cleavage Drilled Compression) system~\cite{Janssen74}.

\end{enumerate}

\begin{figure}
\centering
\includegraphics[height=3.5cm]{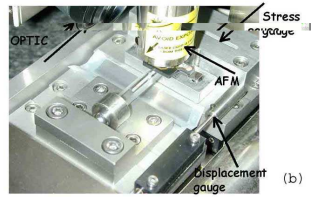}
\caption{Experimental setup.} 
\label{fig1}
\end{figure}

The experimental set-up is illustrated in Fig.~\ref{fig1}. All the experiments are performed at a constant temperature of $22.0\pm 0.5^\circ\mathrm{C}$ in a leak-proof chamber under an atmosphere composed of pure nitrogen and water vapour at a relative humidity level of $42\pm 1\%$ after preliminary out-gassing. 

\begin{figure}
\centering
\includegraphics[width=0.7\textwidth]{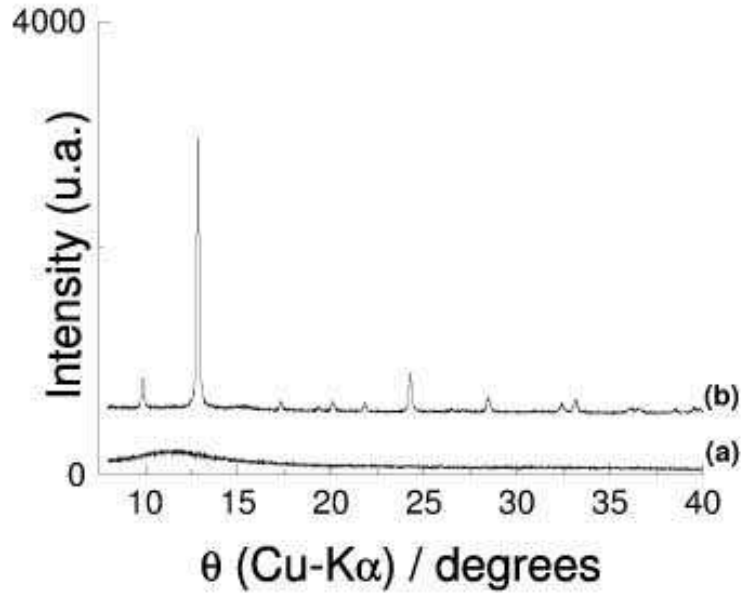}
\caption{ X-ray diffraction patterns on lithium alumino-silicate glass-ceramics. (a): Sample A in a pure glassy state, b): Sample B in a slightly unglassy state.} 
\label{fig2}
\end{figure}

Lithium alumino-silicate glass-ceramics were studied. The main advantage of this kind of materials
is that its structure can be controlled through an applied thermal treatment~\cite{Marliere03}. Two types of structures are investigated below. The first one (sample A) corresponds to a pure glassy state as identified by X-ray diffraction (Fig.~\ref{fig2}a). A single thermal treatment at $660^\circ\mathrm{C}$ is performed in order to remove residual stresses. The second one (sample B) is classified as a slightly unglassy state in which small crystals of $\beta $-quartz phase have formed (Fig.~\ref{fig2}b) in the glassy phase. The structure of sample B is obtained after a two-step thermal treatment at plateau temperatures of $T_{1}$=$750^\circ\mathrm{C}$  and $T_{2}$=$900^\circ\mathrm{C}$. Both the size and the volume fraction of crystalline grains are evaluated by imaging the samples surface by AFM after a stay in fluorhydric acid (concentration ranging from $0.4\%$ to $2\%$) for $30\un{s}$, which dissolves the amorphous phase faster than the crystalline one~\cite{Dickele02}.

\begin{figure}
\centering
\includegraphics[height=3.5cm]{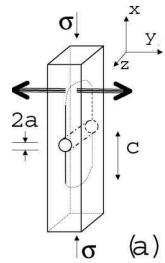}
\caption{Sketch of the DCDC geometry.} 
\label{fig3}
\end{figure}

Fractures are then performed on the DCDC set-up~\cite{Janssen74,He95}: Parallelepipedic ($4\times4\times40\un{mm}^3$) samples (Fig.~\ref{fig3}) are designed with a cylindrical hole (radius $a=0.5\un{mm}$) drilled in the centre and perpendicularly to the $4\times40\un{mm}^2$ surface (Fig.~\ref{fig3}). The hole axis defines the $z$-direction. The $x$-axis (resp. $y$-axis) is parallel to the $40\un{mm}$ (resp. $4\un{mm}$) side of the $4\times40\un{mm}^2$ surfaces.

In all cases, the $4\times40\un{mm}^2$ surfaces are optically polished (the measured RMS roughness is around $0.25\un{nm}$ for a $10\times10~\mu\mathrm{m}^2$ scan size). A compressive load is applied perpendicularly to the $4\times4\un{mm}^2$ surfaces. The external stress $\sigma$ is gradually increased by the slow constant displacement ($0.02\un{mm/min}$) of the jaws of the compressive machine (Fig.~\ref{fig1}). Once the two cracks are initiated -~symmetrically to the hole axis~-, the jaws  displacement is stopped. The crack then propagates along the $x$-axis in the symmetry plane of the sample parallel to the ($x$,$z$) plane. In this geometry, the stress intensity factor $K_I$ is given by~\cite{He95}: $K_I=\sigma\sqrt{a}/(0.375c/a+2)$, where $c$ is the crack length (Fig.~\ref{fig3}).

The crack motion within the ($x$,$y$) surface is monitored by our experimental system combining optical microscopy and AFM. Optical image processing gives the position of the crack tip (with a spatial resolution of 5 micrometers) and consequently the velocity for $v$ ranging from $10^{-6}$ to $10^{-9}\un{m.s}^{-1}$. Let us note that the measured "instantaneous" velocity is  actually averaged over a temporal window the width of which is given by time taken by the crack tip to advance on a distance equal to the spatial resolution (typically 5000s for a mean speed of $10^{-9}\un{m.s}^{-1}$). By AFM measurements~\cite{Marliere01}  -~performed in a high amplitude resonant mode ("tapping" mode)~-, one probes the crack tip neighbourhood at magnifications ranging from $75\times75\un{nm}^2$ to $5\times5~\mu\mathrm{m}^2$. The crack tip can be tracked at velocities ranging from $10^{-9}$ to $10^{-12}\un{m.s}^{-1}$. In this latter case, the minimum width of the temporal window is reduced by two orders of magnitude thanks to the increased spatial resolution. 

\section{Crack tip velocity}

At the very first moments, the crack propagates very quickly. In this regime, the crack velocity $v$ is independent of the chemical composition of the surrounding environment~\cite{Wiederhorn67}. As the crack length $c$ increases, $K_I$ decreases, and $v$ decreases quickly. Under vacuum, the crack stops for $K_I$ smaller than a critical value $K_{Ic}$ referred to as the toughness of the material. But in a humid atmosphere, the corrosive action of water on glass allows for slow crack propagation at much lower values of the stress~\cite{Wiederhorn67}. When $K_I$ becomes smaller than the fracture toughness $K_{Ic}$ (i.e. in the stress corrosion regime), the crack motion is slow enough to be monitored by our experimental system. 

\begin{figure}
\centering
\includegraphics[height=6cm]{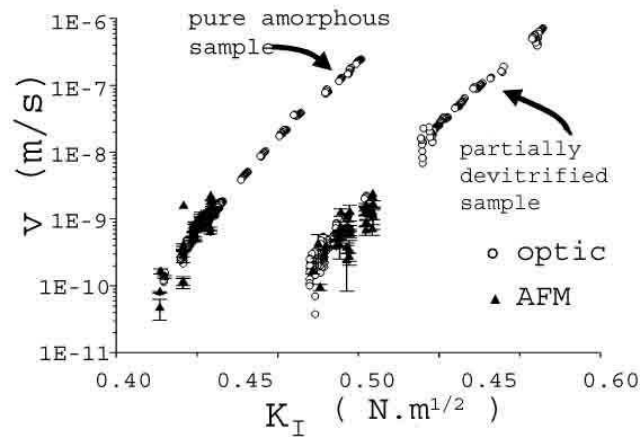}
\caption{Variation of the crack tip velocity $v$ versus the stress intensity factor $K_I$. Open circles (resp.  black triangles) correspond to optical measurements (resp. AFM measurements). The error bars correspond to the standard deviation on the velocity for a fixed value of the stress intensity factor $K_I$. For the lowest velocities $v$, the fluctuations are of the order of the average velocity.} 
\label{fig4}
\end{figure}

Figure~\ref{fig4} shows the variation of the velocity $v$ as a function of the stress intensity factor $K_I$ for both the amorphous and the partially devitrified specimens. The observed exponential behaviour is compatible with a stress enhanced chemical activated process.\cite{Wiederhorn67,Wiederhorn70}.

\begin{figure}
\centering
\includegraphics[height=3.5cm]{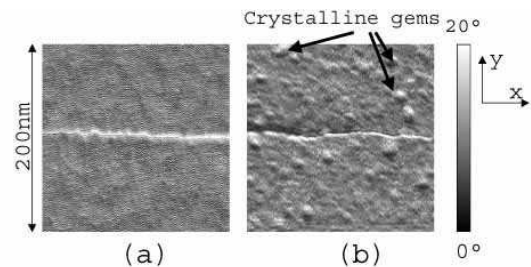}
\caption{Crack path in (a) sample A (the amorphous specimen) and (b) sample B (the partially devitrified one). Frames are phase images sensitive to local mechanical properties, which allows to distinguish the crystalline heterogeneities. In the partially devitrified sample, the crystals deflect the crack which propagates within the amorphous phase.} 
\label{fig5}
\end{figure}

The velocities measured for the partially devitrified sample (Fig.~\ref{fig4}b) are shifted toward positive $K_I$ when compared to the data relative to the amorphous sample. To understand this shift, we probe~\cite{Dickele02} the surface crack path at sub-micrometric scales (Fig.~\ref{fig5}). In the partially devitrified specimen, the crystalline germs deflect the crack (Fig.~\ref{fig5}b). As the crack tip keeps going through the amorphous phase, the stress enhanced chemical activated process is identical in both samples (A and B) and consequently the slope of the semilogarithmic curves $v(K_I)$ are the same in both the amorphous and the partially devitrified specimens. However, the deflections of the crack by the crystals induce local mode II and mode III components in the local stress intensity factor, which toughens the material~\cite{Faber83}. Consequently, $K_{Ic}$ is larger in the devitrified sample, which shifts the $v(K_I)$ curve. A similar effect of the size of heterogeneities has been observed for the fatigue of metallic alloys \cite{Ducourthial01}.

The lowest velocities - only reachable thanks to very high spatial resolution of AFM- shown in Fig.~\ref{fig4} exhibit important fluctuations -~of the order of the average velocity~- for {\em both} specimens~\cite{Celarie03a}. Consequently, they cannot be related to the crystalline heterogeneities but are more likely inherent to the amorphous phase. The velocity fluctuations shown in figure 4 are of the order of $5.10^{-11}\un{m/s}$, {\em i.e.} of the same order of the magnitude as the lowest measured averaged velocities : this fluctuations are the same in both specimens. 

\section{Evidence of nanometric cavities ahead of the crack tip}

\begin{figure}
\centering
\includegraphics[width=0.7\textwidth]{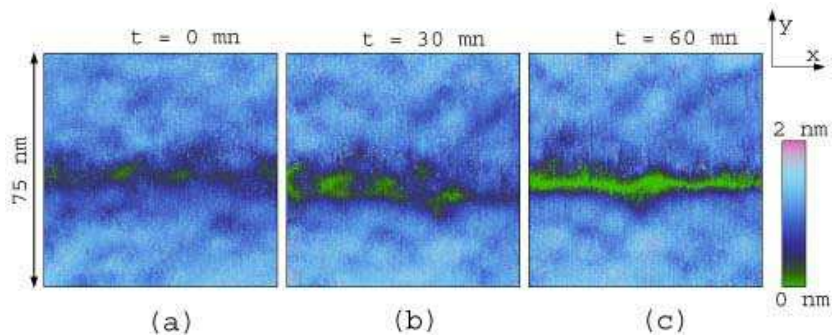}
\caption{Sequence of successive topographic AFM frames showing the crack propagation at the surface of the specimen. The scan-size is $75\times75\un{nm}^2$ and the heights range over $2\un{nm}$. The recording time for one frame is around $3\un{mn}$ and two successive frames are separated by $20\un{mn}$. The crack front propagates from the left to the right ($x$-direction) with a mean velocity $v$ of around $10^{-11}\un{m/s}$. (a): evidence of nanometric damage cavities before the fracture advance. (b): growth of the cavities. (c): the crack is advancing via the coalescence of all the cavities} 
\label{fig6}
\end{figure}

To understand the origin of the velocity fluctuation, we probed the neighbourhood of the crack tip at the nanometer scale in the amorphous specimen (Fig.~\ref{fig6}). This clearly reveals the presence of cavities of typically $20\un{nm}$ in length and $5\un{nm}$ in width ahead of the crack tip (Fig.~\ref{fig6}a)~\cite{Celarie03b}. 

These cavities grow with time (Fig.~\ref{fig6}b) until they coalesce (Fig.~\ref{fig6}c). At these nanometric scales, the crack front does not propagate regularly, but intermittently through the merging of the nano-scale cavities, which explains the large fluctuations observed for the lowest velocitites (Fig.~\ref{fig4}).

\section{Nanometric cavities and nano-ductility.}

\begin{figure}
\centering
\includegraphics[ width=0.7\textwidth]{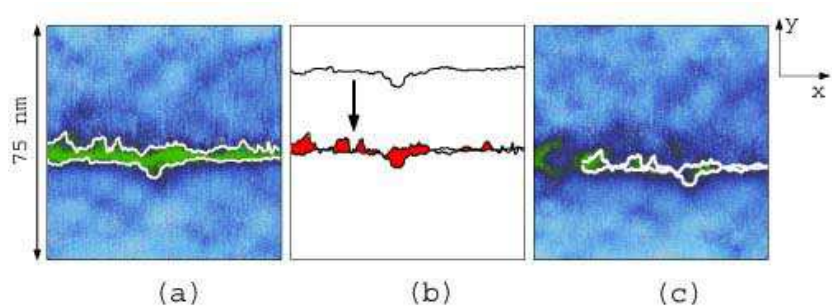}
\caption{Fracture Surface Topographical Analysis (FRASTA). (a) Frame~\ref{fig4}c (broken sample) is binarised and the contours of the crack are determined. (b) The lower line is first numerically raised over the upper one and then gradually displaced in the direction of decreasing $y$, as schematised by the arrow. Cavities are coloured in green. (c): Result of the method: superimposition of the obtained cavities on the image~\ref{fig6}b recorded prior to complete failure.} 
\label{fig7}
\end{figure}

To ensure that the spots observed ahead of the crack tip are actually damage cavities which grow further and coalesce with the main crack leading to failure, we use the Fracture Surface Topography Analysis (FRASTA) technique first introduced to study damage in metallic alloys~\cite{Kobayashi87,Miyamoto90}. These authors showed that, in a ductile scenario, each cavity initiation is accompanied by local irreversible plastic deformations {\em printed} in relief on the developing fracture surfaces (the crack lines when the method is applied in two dimensions as in the present case) that should remain visible after the cavities have coalesced and the crack has crossed. The FRASTA method is designed to analyse the mismatch between the lower fracture surface (the lower crack line when the method is applied in two dimensions as in the present case) and the upper one (the upper crack line), mismatch due to the presence of cavities. For that purpose, the crack lines are first determined by binarising the image of the sample after fracture (Fig.~\ref{fig7}a), and the unbroken material is reconstituted virtually by placing numerically the lower line over the upper one (Fig.~\ref{fig7}b). By translating the lower crack line gradually in the direction of decreasing $y$ (Fig.~\ref{fig7}b), -~this displacement is what actually occurs during fracture~-, one can see the cavities appearing and growing in the chronological order. The structure obtained for a given displacement, \ie at a given time, is superimposed on images recorded prior to failure and shown to correspond actually to cavities observed at this given time (Fig.~\ref{fig7}c). This clearly indicates that the spots shape is determined well ahead of the crack tip, before the first ones can be actually observed, which provides a rather strong argument to relate these nano-scale spots to damage cavities.

\begin{figure}
\centering
\includegraphics[ width=0.7\textwidth]{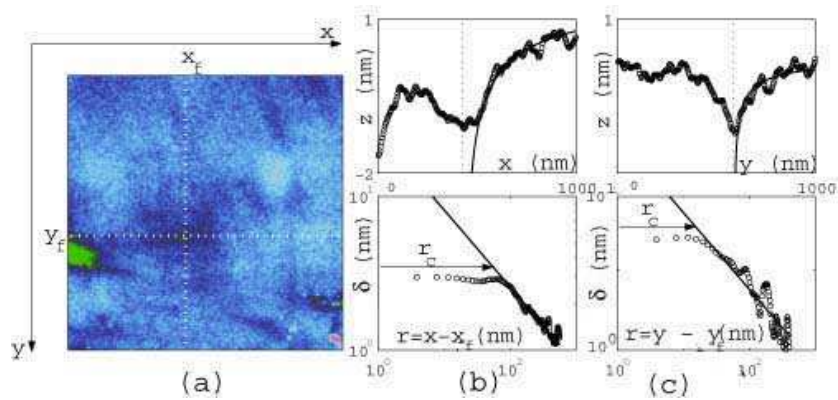}
\caption{Measurements of the surface deformations and comparison with the predictions for an ideal Hookean material. The crack propagates from left to right ($x$ positive). (a): typical AFM topographical frame of the vicinity of the crack tip. The scan-size is $1\times1~\mu\mathrm{m}^2$ and the heights range over $3\un{nm}$. The white vertical (respectively horizontal) dotted line sets the $x$-coordinate $x_f$ (respectively the $y$-coordinate $y_f$) of the crack tip. (b) (respectively (c)) top: Plot of the $z$ profile along (respectively perpendicularly to) the direction of crack propagation. The open circles correspond to experimental data while the full line corresponds to the prediction $z=z_0-Ar^{-1/2}$ (where $z_0$ and $A$ are fit parameters) given for an ideal Hookean solid. Bottom: Log-log plot of the depth $\delta=z_0-z$ versus the distance $r=x-x_f$ (respectively $r=y-y_f$) from the crack tip. For $r\leq r_c$, the $\delta$-profile departs from the predictions given by the linear elastic theory.} 
\label{fig8}
\end{figure}

This nano-scale ductility is also confirmed by the study of the displacement field around the crack tip (see also references~\cite{Guilloteau96,Henaux00} for related discussion): for a slit-like plane crack in an ideal Hookean continuum solid, all the components of the stress tensor are predicted to decrease as $r^{-1/2}$ with the distance $r$ from the crack tip~\cite{Irwin57}. If the macroscopic stress-strain relation work on the nanometer scale the depth $\delta$ (see the caption of Fig.~\ref{fig8} for a precise definition) should thus decrease as $r^{-1/2}$ in the vicinity of the crack tip during crack propagation. To be more precise, $\delta=Ar^{-1/2}\cos(\theta/2)$ for mode I fractures~\cite{Lawn93} where $A$ is a pre-factor depending only on the applied load and on the specimen geometry, and $\theta$ is the angle between the direction of crack propagation and the $\vec{r}$ direction. Measurements of depth profiles have been performed on $1\times1~\mu\mathrm{m}^2$ AFM topographical frames (Fig.~\ref{fig8}a) along the direction of crack propagation (Fig.~\ref{fig8}b) and perpendicularly to it (Fig.~\ref{fig8}c). For both profiles, $\delta$ departs from the linear elastic $r^{-1/2}$ scaling for $r$ smaller than a threshold $r_c$ highly dependent on $\theta$: for $\theta=0^\circ$, $r_c=100\un{nm}$ while for $\theta=90^\circ$, $r_c=20\un{nm}$. These short range departures from the linear elastic behaviour may be related to the presence of cavities although other phenomena could be responsible for this discrepancy (as for instance viscous effect, chemical effect, ...). However, the fact that the order of magnitude of the ratio $r_c(\theta=0^\circ)/r_c(\theta=90^\circ)$ -~much higher than the ratio of the cosine terms in the linear elastic expression of $\delta$~- is close to the aspect ratio of the observed damage cavities, strongly suggests a correlation between damage and non linear elasticity.

\section{Discussion}

Same fracture experiments performed in amorphous silica specimens reveal similar damage cavities. This suggests that their existence does not depend on the precise chemical composition of the studied glass. The origin of the nucleation of cavities should be found more likely in the amorphous structure, which contains inherent atomic density fluctuations at the nanometer scale. Such a scenario was indeed predicted by  Molecular Dynamics Simulation~\cite{Nakano99,vanBrutzel97,vanBrutzel02,Rountree02} that evidenced atomic density fluctuations in the structure of simulated amorphous silica : The Si and O atoms are shown to form silica tetrahedra connected together to build rings of different sizes ranging from 3 to 9 tetrahedra. At larger length scales, ranging from $1.5\un{nm}$ to $6\un{nm}$, the density of these rings is found to fluctuate with high density areas surrounded by low density areas. Moreover, the Molecular Dynamics simulations of van Brutzel~\cite{vanBrutzel97,vanBrutzel02,Rountree02} show that, at this length scale, crack propagates by growth and coalescence of small cavities which appear in areas with low density of rings, ahead of the crack tip. They behave as stress concentrators and grow under the stress imposed by the presence of the main crack to give birth to the cavities actually observed in the AFM frames. 
 
The comparison of these theoretical results and the experimental evidences reported here strongly suggests that the crack advance in glass occurs by a ductile-like process -~at a temperature far below $T_g$~- bound to nucleation, growth and coalescence of damage cavities. The nucleation of these damage cavities is initiated very likely within the low density {\em nanometric} areas. The origin of the universality in the existence of two roughness regimes observed on post-mortem glass fracture surfaces~\cite{Bouchaud97,Daguier95} can now be explained: The fracture advances through the nucleation, growth and coalescence of damage cavities in {\em both} brittle and ductile material. The only difference resides in the length scales of the damage cavities : those related to  'brittle' materials are approximately three orders of magnitude smaller than those relative to metallic alloys.

These observations strongly support the scenario proposed  by Bouchaud {\em et al.}~\cite{Bouchaud02}: The coalescence of the cavities is likely responsible for the observed roughness exponent of the $\zeta =0.8$ regime observed at larger length scales (from $\xi_c$ to $100\un{nm}$). The order of magnitude of $\xi_c$ is indeed in good agreement with our present observations, where cavities at coalescence are a few tens of nanometers wide. On the other side, the $\zeta =0.5$ regime observed for smaller length scales is likely related to an irregular start-stop motion characteristic of the non equilibrium dynamic critical-point at which the crack starts to advance. This may generate diffusely damped crack front corrugations naturally leading to a steady state roughening of fracture surfaces. This effect is predicted to occur even though the macroscopic velocity of the crack front is small provided that the instantaneous velocity during an `avalanche'
in a strongly heterogeneous medium may reach much higher values. This may have been at the origin of the  important fluctuations in crack velocities reported in the present paper.

Here, let us note that AFM observations are performed on the sample surface, where the mechanical state is different from that of the bulk. Hence, the observed sizes and growth rates of cavities at the surface may well differ from those in the bulk. New experiments using the FRASTA method in three dimensions applied to the post-mortem study of the fracture surfaces are currently being performed, in order to have access to the three dimensional structure of bulk damage and its evolution. Through this new set of experiments, one should be able to correlate also the damage structure to the fracture surface morphology.

The topographic study of post-mortem surfaces will also be a good experimental test of the validity of Bouchaud {\em et al.}~\cite{Bouchaud02}. This scenario suggests that small length scale exponent $\zeta =0.5$ is due to the existence of diffusively broadened crack front waves in localised depinning events. Other important predictions are that roughness amplitude should be strongly anisotropic and this effect should be very pronounced at low crack growth velocities. That model also predicts that the physics of the formation and coalescence of the cavities is responsible for the observed roughness exponent of $\zeta = 0.8$. Such a scenario was already proved to be realistic thanks  to observations of the roughness of growing cavities before coalescence in an aluminium alloy~\cite{Paun}. It will be of course crucial -even if much more experimentally difficult due to much lower $\xi_c$ value- to check these hypotheses in the case of nano-ductile materials.

Moreover, the structure of damage, which influences macroscopic mechanical properties such as fracture toughness and lifetime, should then be linked to the glass composition and nanostructure~\cite{vanBrutzel97,Falk99,Benoit00,Jund01}. Complementary analyses addressing the question of the chemical bonds on the fracture surface will also be performed.

\section{Conclusion}

Quasi-static fracture in glassy materials has been studied {\em in real time} at sub-micrometric length scales in a wide velocity range. The crack velocities are measured as a function of the stress intensity factor in both amorphous materials and a partially devitrified one. Important fluctuations are reported for the lowest velocities. They are conjectured to be related to the presence -~inside of the glassy phase~- of nano-scale damage cavities observed in real time ahead of the crack tip. The influence of these damage cavities on mechanical properties in the non purely elastic zone in the vicinity of the crack tip have been presented. These results confirm the scenario proposed by Bouchaud {\em et al.}~\cite{Bouchaud02} to explain the origin of the two self-affine regimes observed on fracture surfaces. The implications of such a ductile fracture mode in glass on the morphology of the fracture surfaces have been discussed. However it should be noted that our investigations are performed on the sample surface while the fracture surface morphology is related to the fracture of bulk. Consequently, it would be interesting to investigate the 3D distribution of damage cavities. Work in this direction is currently in progress.

Finally, the fact that glass, at temperatures far below the glass transition temperature $T_g$, joins the class of damageable materials should have important consequences for its mechanical properties. In applications, the design of structures using glass might be modified to take this behaviour into account, especially for slow crack propagation processes.

The similarity between the damage modes of materials as different as glass and metallic alloys is an important clue to understand the origin of puzzling universal behaviors, hence shedding new light into the basic physical mechanisms of fracture.


\end{document}